\documentclass[conference]{IEEEtran}
\usepackage[latin9]{inputenc}
\usepackage{amsmath}
\usepackage{amssymb}
\usepackage{graphicx}

\makeatletter

\providecommand{\tabularnewline}{\\}
\newcommand{\lyxdot}{.}

\IEEEoverridecommandlockouts
\usepackage{cite}
\usepackage{amsfonts}\usepackage{algorithmic}
\usepackage{textcomp}
\usepackage{xcolor}
\usepackage{tikz}
\usepackage{hyperref}
\usepackage{lipsum}
\def\BibTeX{{\rm B\kern-.05em{\sc i\kern-.025em b}\kern-.08em
    T\kern-.1667em\lower.7ex\hbox{E}\kern-.125emX}}

\@ifundefined{showcaptionsetup}{}{%
 \PassOptionsToPackage{caption=false}{subfig}}
\usepackage{subfig}
\makeatother

\newcommand\copyrighttext{%
	\footnotesize \textcopyright 2021 IEEE. Personal use of this material is permitted.
	Permission from IEEE must be obtained for all other uses, in any current or future
	media, including reprinting/republishing this material for advertising or promotional
	purposes, creating new collective works, for resale or redistribution to servers or
	lists, or reuse of any copyrighted component of this work in other works.
	DOI: \href{https://doi.org/10.1109/CCNC49033.2022.9700718}{10.1109/CCNC49033.2022.9700718}}
\newcommand\copyrightnotice{%
	\begin{tikzpicture}[remember picture,overlay]
		\node[anchor=south,yshift=10pt] at (current page.south) {\fbox{\parbox{\dimexpr\textwidth-\fboxsep-\fboxrule\relax}{\copyrighttext}}};
	\end{tikzpicture}%
}

\begin{document}
\title{Polar Coding for Efficient Transport Layer Multicast}
\author{\IEEEauthorblockN{Pablo Gil Pereira and Thorsten Herfet, \IEEEmembership{IEEE Senior Member}}
\IEEEauthorblockA{\textit{Telecommunications Lab}\\
\textit{Saarland Informatics Campus, D-66123}\\
Saarbrücken, Germany\\
\{gilpereira, herfet\}@cs.uni-saarland.de}}
\maketitle
\copyrightnotice
\begin{abstract}
In this paper, we shed light on how an adaptive, efficient error coding
in the transport layer helps ensure the application's requirements.
We recap the use of MDS codes and show that binary coding can significantly
reduce the complexity and hence increase the applicability also for
embedded devices. We exploit the persymmetric structure of the generator
matrix in polar codes to establish a duality of dispersion over channels
(the polarization effect) and over packets (the generality required
for multicast transmission), thereby constructing systematic polar
codes for incremental redundancy whose performance, despite a much
lower complexity, is near to MDS codes for medium-range residual loss
rates.
\end{abstract}


\section{Introduction}

Fog computing brings large computational resources closer to the end
devices, thereby enabling applications demanding low latency~\cite{bonomi2012fog}.
Networked Control Systems (NCSs) are a paradigmatic example of these
applications as they require predictably low latency, as well as reliability~\cite{lee2008cyber}.
The transport layer is the only one in the protocol stack with an
end-to-end perspective of the communication and thus it is fundamental
to achieve timeliness and reliability guarantees~\cite{saltzer1984end}.
This layer can monitor the lower layers and adapt to them to ensure
the quality of service by means of a time-aware error control function.
The most common error control schemes, namely Automatic Repeat reQuest
(ARQ) and Forward Error Coding (FEC), have inherently different timing
characteristics---i.e., one is reactive whereas the other is proactive.
Hybrid ARQ (HARQ) can combine them in order to achieve optimal performance
in information-theoretical terms under timing constraints~\cite{huang2009hybrid,gorius2012predictably}.

Maximum Distance Separable (MDS) codes have traditionally been used
at the transport layer for FEC~\cite{rizzo1997effective,gorius2012predictably,palmer2018quic,michel2019quic}
because the number of correctable losses equals the number of transmitted
parity packets. However, MDS codes perform computationally expensive
operations in high order Galois Fields, resulting in unacceptable
delay and energy consumption in embedded devices. Binary codes~\cite{luby2002lt,mackay2005fountain,shokrollahi2004ldpc}
have been designed to reduce the coding complexity, but they have
the disadvantage of requiring an excess of parity packets and do not
perform well in multicast scenarios. In this paper, we analyze the
recently developed polar codes~\cite{arikan2009channel} and exploit
the persymmetric structure of their generator matrix to construct
binary codes that are more efficient than previous binary codes for
the transport layer while approaching the performance of MDS in multicast
at the same time. The contribution of this paper is threefold:
\begin{itemize}
\item To the best of our knowledge, this paper provides the first analysis
of polar codes at the transport layer from a joint energy and multicast
perspective.
\item A code construction mechanism is proposed that exploits polar codes'
dual dispersion over channel and receivers to enable binary codes
in multicast.
\item We show that the proposed code approaches MDS codes for multicast
with much lower complexity, while the excess packets are kept low
for short block lengths.
\end{itemize}
The remainder of the paper is organized as follows: Section~\ref{sec:Related-work}
presents related work and some background is discussed in Section~\ref{sec:Background}.
The impact of the transport layer in HARQ is analyzed in Section~\ref{sec:complexity-dilemma}.
Section~\ref{sec:Energy-efficient,-Multicast-HARQ} discusses energy
and multicast aspects of error coding and Section~\ref{subsec:Multicast-Polar-Codes}
describes the proposed code construction, which is then evaluated
in Section~\ref{sec:Evaluation}. Section~\ref{sec:Conclusion}
concludes the paper.

\section{Related Work\label{sec:Related-work}}

There have been many proposals to complement lower layer error coding
with coding at the transport layer in order to improve reliability
without prohibitively increasing the delay. MDS codes have been used
to provide predictable reliability under time constraints~\cite{gorius2012predictably},
improve video streaming quality~\cite{palmer2018quic} and mitigate
feedback implosion in multicast~\cite{rubenstein2001study}. Windowed
random linear codes (RLC) are an alternative to block codes that make
the end-to-end delay independent of the block length by evenly spreading
the parity packets over the source packets. Huang et al.~\cite{huang2009hybrid}
implement RLC in the transport layer together with ARQ, Roca et al.~\cite{roca2017less}
show that RLC reduces the FEC delay at the expense of lower code rates
than block codes and Karzand et al.~\cite{karzand2017design} use
it to reduce the delay in in-order delivery. Although we have opted
to use block codes that in principle induce a larger delay, our proposed
error coding is delay-aware, which ensures that the delay does not
exceed the application requirements. Other approaches have focused
on simplicity and solely apply exclusive or (xor) to all the packets
in the block~\cite{garrido2019rquic,QuicFecv1,ferlin2018mptcp},
thereby allowing the recovery of a single loss. Those codes are a
subset of the proposed polar coding in this paper because the parity
transmission always starts with the xor of all packets. Michel et
al.~\cite{michel2019quic} implemented the three aforementioned approaches
in QUIC and show that RLC achieves the best performance in terms of
delay.

\section{Background\label{sec:Background}}

\begin{figure}
\centering\scalebox{0.75}{
	\begin{tikzpicture}

		\draw[-] (0,-0.3) -- (0,-2.5);
		\draw[-] (-0.05,-2.36) -- (0.05,-2.46);
		\draw[densely dotted] (0,-2.5) -- (0,-2.8);
		\draw[-] (-0.05,-2.64) -- (0.05,-2.74);
		\draw[->] (0,-2.7) -- (0,-7.35);

		\draw[-] (4,-0.3) -- (4,-2.5) ;
		\draw[-] (3.95,-2.36) -- (4.05,-2.46);
		\draw[densely dotted] (4,-2.5) -- (4,-2.8);
		\draw[-] (3.95,-2.64) -- (4.05,-2.74);
		\draw[->] (4,-2.7) -- (4,-7.35);

		\node (t) at (-0.5,-7.35) {$t$};

		\draw (0,-0.5) rectangle (-0.5,-1);
		\draw[->] (0,-0.5)-- (4,-0.85);
		\draw[dashed] (0,-1) -- (4,-1.35);
		\node at (-0.25,-0.75) {\small $m_{1}$};

		\draw (0,-1.4) rectangle (-0.5,-1.9);
		\draw[->] (0,-1.4)-- (4,-1.75);
		\draw[dashed] (0,-1.9) -- (4,-2.25);
		\node at (-0.25,-1.65) {\small $m_{2}$};

		\draw (0,-2.9) rectangle (-0.5,-3.4);
		\draw[->] (0,-2.9)-- (4,-3.25);
		\draw[dashed] (0,-3.4) -- (4,-3.75);
		\node at (-0.25,-3.15) {\small $m_{k}$};

		\draw (0,-4) rectangle (-0.5,-4.5);
		\draw[->] (0,-4) -- (4,-4.35);
		\draw[dashed] (0,-4.5) -- (4,-4.85);
		\draw[->] (4,-4.85) -- (0,-5.2);
		\node at (-0.25,-4.25) {\small $p_{1}$};
		
		\draw (0,-5.2) rectangle (-0.5,-5.7);
		\draw[->] (0,-5.2) -- (4,-5.55);
		\node at (-0.25,-5.45) {\small $p_{2}$};
		
		\draw (0,-5.7) rectangle (-0.5,-6.2);
		\draw[->] (0,-5.7) -- (4,-6.05);
		\draw[dashed] (0,-6.2) -- (4,-6.55);
		\node at (-0.25,-5.95) {\small $p_{3}$};

		\draw[-] (4.5,-0.5) -- (4.5,-7.15);
		\draw[-] (4.45,-0.5) -- (4.55,-0.5);
		\node at (5,-0.65) {\small $\frac{RTT}{2}$};
		\draw[-] (4.45,-0.85) -- (4.55,-0.85);
		\node at (5,-2.45) {\small $k\cdot T_{s}$};
		\draw[-] (4.45,-3.75) -- (4.55,-3.75);
		\node at (5,-4.15) {\small $D_{enc}$};
		\draw[-] (4.45,-4.35) -- (4.55,-4.35);
		\node at (5,-4.75) {\small $D_{tx}$};
		\draw[-] (4.45,-4.85) -- (4.55,-4.85);
		\node at (5,-5.2) {\small $RTT$};
		\draw[-] (4.45,-5.55) -- (4.55,-5.55);
		\node at (5.15,-6.05) {\small $2 \cdot D_{tx}$};
		\draw[-] (4.45,-6.55) -- (4.55,-6.55);
		\node at (5,-6.85) {\small $D_{dec}$};
		\draw[-] (4.45,-7.15) -- (4.55,-7.15);
	\end{tikzpicture}
}\caption{HARQ delay budget. It considers the RTT, IPT ($T_{s}$), transmission
delay ($D_{tx}$), block length ($k$), and encoding ($D_{enc}$)
and decoding delay ($D_{dec}$).\label{fig:delay-budget}}
\end{figure}
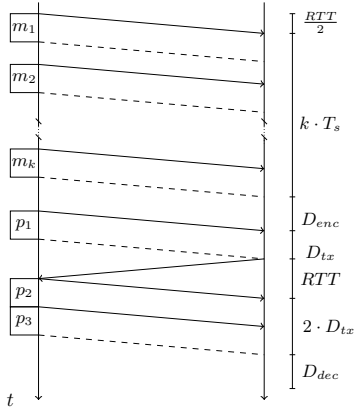

Once the application requirements are known, a time-aware error control
could adapt its performance to meet them with a minimal resource footprint.
Monitoring the residual loss rate and delay of the lower layers, the
transport layer can react only when strictly necessary to guarantee
the end-to-end performance---i.e., only add redundancy if the lower
layers do not provide enough reliability. This adaptability can be
achieved by dynamically changing the partitioning between ARQ, whose
delay is dominated by the round-trip time (RTT) as packets are retransmitted
upon the detection of losses with some acknowledgement mechanism,
and FEC, whose delay is dominated by the inter-packet time (IPT) that
is required to collect packets before encoding. ARQ only retransmits
redundancy when required, thereby increasing efficiency. On the other
hand, FEC is a better alternative under tight delay constraints as
the sender does not need to wait for feedback, as well as in multicast,
where collecting feedback from all of the receivers is infeasible~\cite{delucia1997multicast}
and the generality of repair packets is important. Therefore, under
delay constraints, an adaptive HARQ mechanism is optimal in information-theoretical
terms. Figure~\ref{fig:delay-budget} shows the delay budget of a
HARQ scheme. Although the coding process entails complex operations,
the exponential increase in processing power in the last decades made
it possible to run this operation in most machines with negligible
delay. However, this is not the case when the protocol is executed
on embedded devices, resulting in a large delay and energy consumption
that should be minimized.

\section{The complexity dilemma revisited\label{sec:complexity-dilemma}}

Block codes encode a message $m$ with $k$ symbols into a codeword
$c$ via multiplication with a $k\times n$ generator matrix $G$
($c=m\cdot G$). At the receiving end, the original message can be
recovered by solving $m=\hat{c}\cdot\hat{G}^{-1}$, where $\hat{c}$
is a subset of $k$ symbols of $c$ and $\hat{G}$ is a $k\times k$
submatrix of $G$. The encoding process entails a matrix-vector multiplication
with complexity $\mathcal{O}(kn)$, whereas the matrix inversion has
complexity $\mathcal{O}(k^{3})$\footnote{Some decoder optimizations have managed to bring this complexity down
to $\mathcal{\sim O}(k^{2})$ or even $O(n\cdot log(n))$.} and therefore dominates the overall process. In the following sections,
we introduce two types of codes that use different approaches to create
$G$: MDS and binary codes.

\subsection{Maximum Distance Separable Codes}

The principal characteristic of MDS codes is that they fulfil the
Singleton Bound with equality---i.e., $d_{min}=e+1$ where $d_{min}$
is the minimum distance between codewords and $e=n-k$ the number
of correctable erasures~\cite{macwilliams1977theory}. They are defined
on $GF(2^{q})$ where $q$ is the number of bits per symbol. A usual
configuration is $GF(2^{8})$ such that each symbol is a byte. In
order to correct $e$ erasures with any $k$ received packets, every
$k\times k$ submatrix of $G$ should be invertible. From all the
available mechanisms to construct such a matrix~\cite{bloemer1995xor,rizzo1997effective},
this paper focuses on systematic Vandermonde MDS codes, which does
not limit the analysis because the basic operations remain the same.
The particular characteristic of systematic codes is that a verbatim
copy of the original message is present in the codeword. In other
words, the generator matrix has a systematic part, consisting of a
$k\times k$ identity matrix, and a $k\times(n-k)$ parity part ($G=[I|P]$).
Although the code properties are unchanged, systematic codes have
a better residual loss rate as some of the original symbols can still
be recovered even when more than $n-k$ erasures occur. The complexity
of systematic MDS codes is $\mathcal{O}(kp)$ for encoding and $\mathcal{O}(kp^{2})$
for decoding~\cite{rizzo1997effective}---$p$ is the number of
parity symbols---reducing the complexity if $k>>p$, which is usually
the case in multimedia applications, as shown in Section~\ref{sec:Evaluation}.

\subsection{Binary Codes\label{subsec:Binary-Codes}}

Binary codes are defined in $GF(2)$ and they perform encoding and
decoding with simple xors instead of arithmetic operations in higher-order
fields. As a result, they can use graph-based decoding algorithms~\cite{macwilliams1977theory},
thereby reducing the complexity as no explicit matrix inversion is
needed. The complexity of binary codes depends on the average degree
of $G$ and they achieve quasilinear complexity when sparse matrices
are employed. Despite this complexity reduction, using $GF(2)$ does
not guarantee the invertibility of every $k\times k$ submatrix. Consequently,
binary codes require excess symbols to approach the channel capacity,
giving place to the ``\textit{complexity dilemma}'': the computational
complexity is reduced via basic xor operations and sparse matrices,
whereas the network complexity increases due to the redundancy excess.
It can be shown that the excess portion in the transmitted redundancy
decreases when large block lengths are used---e.g., 10.000~\cite{shokrollahi2004ldpc,luby2002lt}.
Although these block lengths are common in the physical layer, they
are infeasible in the transport layer.

\subsection{The Role of the Transport Layer\label{subsec:Role-Transport}}

IP networks ``see'' packetized erasure channels---e.g., erroneous
packets are not forwarded to the upper layers and complete packets
are dropped due to congestion. Therefore, although HARQ at the physical
layer uses bits as symbols, at the transport layer it should recover
complete packets---we assume a 1500-byte MTU throughout the paper.
Consequently, $k$ packets need to be collected before encoding, resulting
in very short codes under time constraints. For example, a video conference
stream at $10\,Mb/s$ with a target delay of $100\,ms$ can only collect
84 packets before the time constraint expires, a difference of 3-4
orders of magnitude with the physical layer. This blocklength difference
makes the matrix inversion no longer dominate the complexity! While
a single matrix inversion is required per block, the encoding step
is iterated throughout the whole packet. Under time constraints, $MTU>>k$
and in networks with small erasure rates $k>>p$. If $\mathcal{C}_{mul}=MTU\times\mathcal{O}(kp)$
is the complexity of the matrix-vector multiplication and $\mathcal{C}_{inv}=\mathcal{\mathcal{O}}(kp^{2})$
the complexity of the matrix inversion, at the transport layer $\mathcal{C}_{mul}>>\mathcal{C}_{inv}$.
Using basic xor, binary codes implement a much simpler matrix-vector
multiplication than MDS codes. However, the most efficient binary
codes use sparse binary matrices that require large block lengths
to reduce the redundancy excess~\cite{luby2002lt,shokrollahi2004ldpc}.
Fully (pseudo-)random fountain codes~\cite{mackay2005fountain} and
polar codes~\cite{arikan2009channel} are an interesting alternative,
as they would benefit from the complexity reduction of binary codes
with a lower transmission overhead due to their non-sparse matrices.

In random fountain codes the parity packets include each source packet
or not with equal probability, resulting in a $\frac{k}{2}$ average
column degree. Therefore, the coding complexity of systematic random
fountain codes is $\mathcal{O}(\frac{k}{2}p)$. On the other hand,
polar codes exploit the polarization effect of binary codes, which
can be seen as a combination of $n$ independent binary channels whose
capacity polarizes to either almost perfect or almost fully noisy.
Polar codes build $n\times n$ generator matrices where the information
symbols are complemented by frozen symbols---typically set to 0---located
in the worst channel positions. If the channels are the columns of
the generator matrix, then the row with the most $1$s disperses the
source symbol over the most channels, thereby achieving the highest
channel capacity. The Bhattacharyya parameter has been proposed to
measure the channel quality~\cite{arikan2009channel}, which for
the binary erasure channels (BEC) considered in this paper exactly
calculates the erasure probability for each channel. The deterministic
generator matrix construction in polar codes allows to i) efficiently
en-/decode with complexity $\mathcal{O}(kp)$, and ii) construct codes
that perform well in multicast, something impossible with fountain
codes due to their randomness.

\section{Energy-efficient, Multicast HARQ\label{sec:Energy-efficient,-Multicast-HARQ}}

The dominant operation in the HARQ complexity at the transport layer
is the matrix-vector multiplication---see Section~\ref{sec:complexity-dilemma}---and
therefore the efforts to reduce the energy consumption should focus
on it. Processing a column in MDS codes entails the multiplication
of $k$ symbols and the results of these multiplications are then
xored. All these operations can be implemented as the summation of
8-bit unsigned integers~\cite{rizzo1997effective}, resulting in
$2k-1$ operations per column. This process is iterated 1500 times
to en-/decode all the bytes in the MTU, although several instructions
can concurrently run when executed on CPUs supporting SIMD instruction
sets. On the other hand, binary codes only require on average $\frac{k-1}{2}$
xors per column, as no multiplications are needed due to the binary
matrix. The binary nature of these codes also allows the grouping
of several bits into a single instruction, thereby reducing the required
instructions to process a packet and hence the delay and energy consumption.
For example, 32-bit CPUs can reduce the required instructions per
packet xoring 32 bits together. Although increasing the field order
results in a similar effect in MDS codes, this comes at the expense
of large memory usage as the complete field is stored in memory to
avoid modulo operations~\cite{rizzo1997effective}.

Unlike MDS codes, binary codes do not include information from all
source packets in each parity packet. This is a consequence of operating
in $GF(2)$, as $G$ includes $0$s to produce linearly independent
columns. Although this reduces the complexity, it is undesirable for
multicast, where the receivers may experience different erasure patterns.
The column degree is a measure of parity packet generalizability:
while the most general column has all bits to one, the degree one
columns are the least general ones. In a multicast deployment, the
average degree of the generator matrix should be maximized---as long
as linearly independent columns can still be generated---and each
new column should maximize the number of correctable erasure patterns.
Therefore, for binary codes to be suitable for energy-efficient, multicast-enabled
HARQ at the transport layer, they should fulfil that i) the number
of excess packets is low for short block lengths, and ii) a mechanism
can be found to maximize the number of satisfied receivers in each
transmission round.

\section{Multicast Polar Codes\label{subsec:Multicast-Polar-Codes}}

It can be observed that the $4\times4$ polar generator matrix $G$
below has a persymmetric structure---i.e., the matrix is symmetric
with respect to the upper-right to the lower-left diagonal. While
the last row corresponds to the largest temporal dispersion, conversely,
the first column is the column with the largest dispersion over receivers
in a multicast group---a.k.a. the most generalizable parity packet.
Therefore, the same Bhattacharyya parameter used to calculate the
reliability of each channel can merely be flipped to measure the quality
of each column for multicast. We use this property to generate incremental
redundancy for multicast. In order to do so, a mechanism is required
that transforms the code into systematic form such that the parity
part $P$ and the original matrix $G$ share those columns with the
lowest Bhattacharyya parameter.

\[
G=\left[\begin{array}{cccc}
1 & 0 & 0 & 0\\
1 & 1 & 0 & 0\\
1 & 0 & 1 & 0\\
1 & 1 & 1 & 1
\end{array}\right]
\]

Assume a vector $\mathcal{A}_{i}$ containing the indices of the $k$
best channels in decreasing quality order and the complementary vector
$\mathcal{A}_{i}^{C}$ with the $n-k$ remaining channels in increasing
quality order. For example, for the $(16,8)$ code in Section~\ref{subsec:Evaluation-Multicast},
the indices of the channels in decreasing quality are $[16,15,14,12,8,13,11,10,7,6,4,9,5,3,2,1]$,
so that $\mathcal{A}_{i}=[16,15,14,12,8,13,11,10]$ and $\mathcal{\mathcal{A}}_{i}^{C}=[1,2,3,5,9,4,6,7]$.
The proposed mechanism to turn $G$ into systematic form is as follows:
the worst channels are used for the frozen bits, meaning that the
rows in $\mathcal{A}_{i}^{C}$ are ignored because they only contribute
a $0$ to the xor. However, instead of using the channels in $\mathcal{A}_{i}$
as usual, the precode turns these channels into the systematic part.
The $k$ columns in $\mathcal{A}_{i}$ are iteratively substituted
by the columns in $I$ the $k\times k$ identity matrix---due to
the fact that $G^{-1}=G$ for polar codes this is the same as pre-multiplying
with $G(\mathcal{A}_{i},\mathcal{A}_{i})$. Finally, the reservoir
of parity packets consists of the columns in $\mathcal{A}_{i}^{C}$.
The persymmetric structure of the generator matrix means that ordering
the channels in decreasing quality equals ordering the columns in
increasing generalizability. Therefore, the parity columns in the
reservoir should be transmitted in the order given by $\mathcal{A}_{i}^{C}$
to maximize the HARQ performance in multicast.

\section{Evaluation\label{sec:Evaluation}}

This section evaluates the proposed polar coding in terms of the required
excess packets, encoding and decoding delay on an embedded platform
and loss recovery in a multicast group.

\subsection{Methodology}

For a fair comparison between the codes, we analyze their performance
when meeting a target packet loss rate ($PLR_{T}$). Assuming an i.i.d.
BEC with erasure probability $p_{e}$, the packet loss rate can be
calculated with Eq.~\ref{eq:bec_plr}.

\begin{equation}
PLR_{BEC}=\frac{1}{k}E\left[I_{k}\right]=\frac{1}{k}\sum_{i=1}^{k}i\cdot P(I_{k}=i)\label{eq:bec_plr}
\end{equation}

$P(I_{k}=i)$ is the probability of $i$ erasures and it depends on
the used code. The expression for MDS codes is given in Eq.~\ref{eq:mds_p},
where $p_{d}\binom{e}{i}$ is the probability that exactly $i$ out
of $e$ erasures in a block fall into the systematic part.

\begin{equation}
P(I_{k}=i)=\sum_{e=max(n-k+1,i)}^{n-k+i}\binom{n}{e}\cdot p_{e}^{e}\cdot(1-p_{e})^{n-e}\cdot p_{d}\binom{e}{i}\label{eq:mds_p}
\end{equation}

\begin{equation}
p_{d}\binom{e}{i}=\frac{\binom{k}{i}\cdot\binom{n-k}{e-i}}{\binom{n}{e}}\label{eq:p_d}
\end{equation}

The same probability for random fountain codes is given in Eq.~\ref{eq:fountain_p},
where $\delta_{e}$ is the probability that the decoding fails after
$e$ erasures. While the decoding always fails when less than $k$
packets are received, fountain codes also have a decoding failure
probability when enough packets are received. We have used the upper
bound to calculate this probability, which converges quickly to the
true value for $k>10$~\cite{mackay2005fountain}.

\begin{equation}
P(I_{k}=i)=\sum_{e=i}^{n-k+i}\delta_{e}\cdot\binom{n}{e}\cdot p_{e}^{e}\cdot(1-p_{e})^{n-e}\cdot p_{d}\binom{e}{i}\label{eq:fountain_p}
\end{equation}

\begin{equation}
\delta_{e}=\begin{cases}
2^{-(n-k-e)} & i\leq e\leq n-k\\
1 & n-k+1\leq e\leq n-k+i
\end{cases}\label{eq:fountain_delta}
\end{equation}

To the best of our knowledge, there is no known analytical expression
$P(I_{k}=i)$ for polar codes. Therefore, we empirically calculated
it by constructing the matrix, emulating $R$ receivers and checking
how many of them could recover from losses. The ground truth has been
calculated for short block lengths to select the number of receivers
$R=50.000$, which achieves an estimation error two orders of magnitude
below the used PLR (see Table~\ref{tab:PLR_error}).

\begin{table}
\begin{centering}
\begin{tabular}{|c|c|c|c|c|c|}
\hline 
Code ($n$,$k$) & (8,4) & (8,6) & (16,8) & (16,10) & (16,12)\tabularnewline
\hline 
\hline 
Error ($\times10^{-5}$) & 4.405 & 4.608 & 4.788 & 2.805 & 4.190\tabularnewline
\hline 
\end{tabular}
\par\end{centering}
~\caption{PLR estimation error for polar coding when using simulated BEC in
comparison to the ground-truth. The BEC has been simulated with $R=50.000$
receivers.}
\label{tab:PLR_error}
\end{table}

In the following, both classes of codes are analyzed considering four
metrics: i) amount of redundancy, ii) encoding and decoding delay,
iii) energy consumption and iv) multicast performance. A Raspberry
Pi Zero W running the Raspbian Buster operating system with Linux
kernel 4.19 has been used as the evaluation platform and the energy
consumed by it has been measured with an LTC2991 board. For the MDS
code, we have used the C implementation in~\cite{rizzo1997effective},
whereas binary codes have been implemented in Rust\footnote{https://git.nt.uni-saarland.de/open-access/ccnc2022}.

\subsection{Redundancy, Delay and Energy Consumption\label{subsec:Redundancy}}

\begin{figure}
\centering{}\includegraphics[scale=0.38]{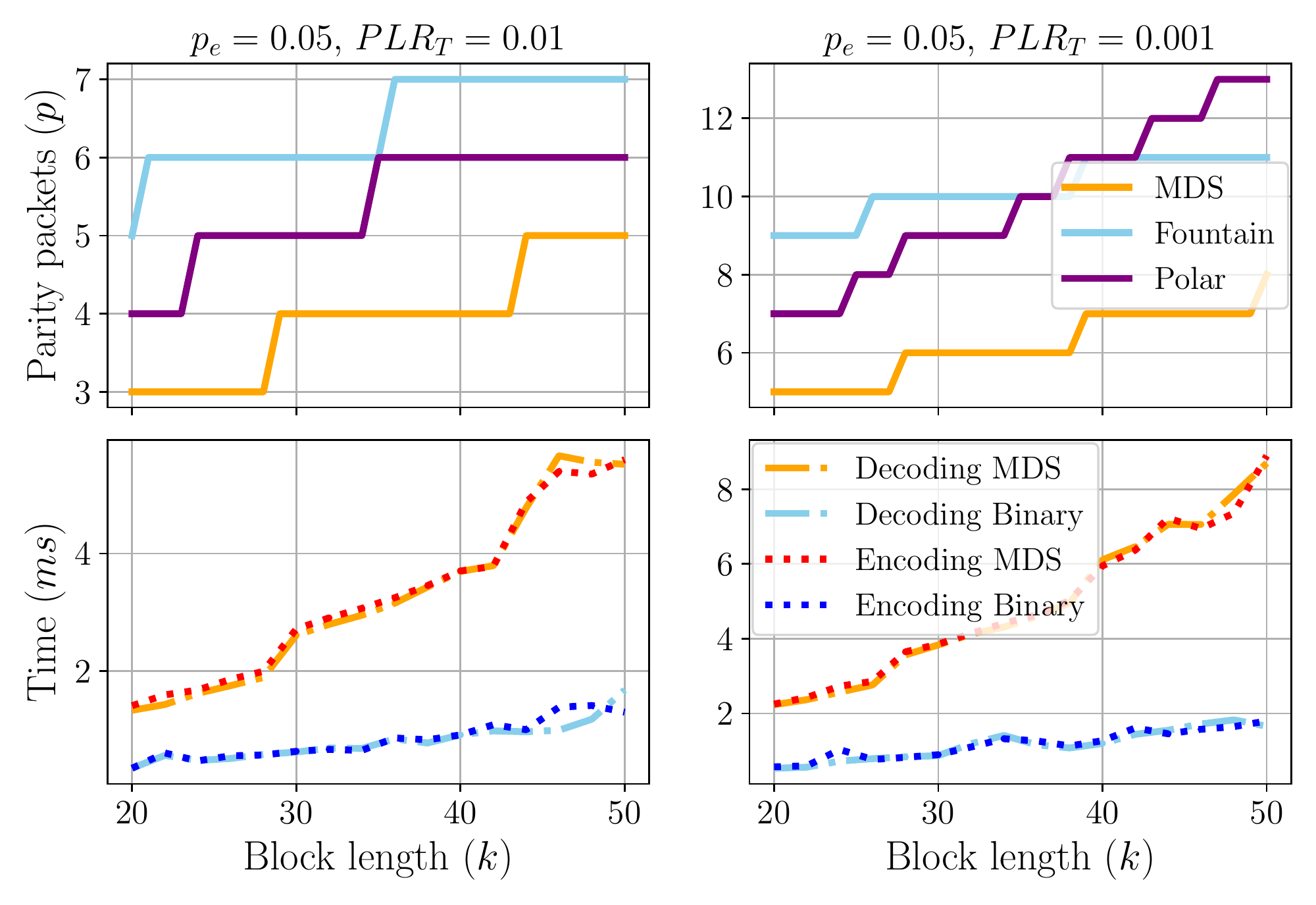}\caption{Required parity packets, encoding delay and decoding delay for MDS,
fountain and polar codes to meet the constraints $PLR_{T}=\{0.01,0.001\}$
in a channel with $p_{e}=0.05$. Binary codes achieve a much lower
delay despite en-/decoding more parity packets. \label{fig:Parity-Delay}}
\end{figure}

For the estimation of the transmission overhead, a WiFi-like deployment
with $p_{e}=0.05$ is considered as well as two different reliability
requirements $PLR_{T}=\{0.01,0.001\}$. Figure~\ref{fig:Parity-Delay}
shows that the proposed polar coding clearly outperforms fountain
codes for short block lengths. Although for $PLR_{T}=0.001$ and $k>38$
there are some cases in which polar codes require the most parity
packets, the validity of the proposed approach still depends on the
ratio between transmission delay, and encoding and decoding delay.
The latter are shown in the lower two plots in Figure~\ref{fig:Parity-Delay},
where both codes correct $e=8$ erasures to consider the worst-case
delay in MDS and perform a fair comparison between them. In this experiment,
only random fountain codes have been considered because, despite achieving
the same $PLR_{T}$, the proposed polar code fails to decode when
the number of erasures is large ($e\geq4$), which also explains its
poor performance for larger block lengths. However, as the underlying
coding operations are the same, fountain codes already show the benefits
of binary codes. For $PLR_{T}=0.01$ the binary code achieves $3.5\times$
lower delay, whereas $5\times$ reduction is achieved for $PLR_{T}=0.001$.
Considering these delay differences, binary codes require at least
$15\:Mb/s$ to compensate such an excess---$7.5\:Mb/s$ when both
ends are embedded devices---, which is small in comparison to the
data rates in 5G and WiFi-6. On top of that, these communication technologies
are more energy-efficient than previous generations~\cite{bjornson2018energy,xu2020understanding},
which emphasizes the relevance of codes that reduce the CPU load.

\begin{figure}
\begin{centering}
\includegraphics[scale=0.35]{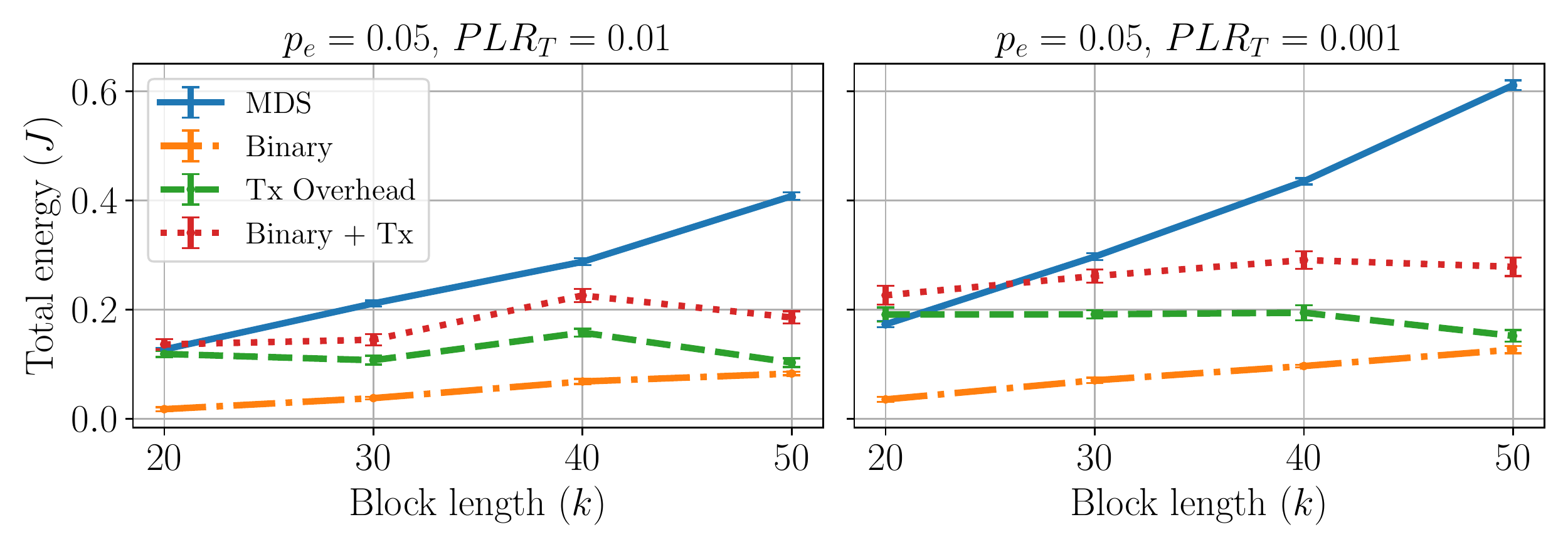}
\par\end{centering}
\centering{}\caption{Total energy consumption for $PLR_{T}=\{0.01,0.001\}$ and $p_{e}=0.05$.
Encoding and decoding are executed 100 times for each code type and
the redundancy excess of binary codes is transmitted 100 times. No
code is optimal for all block lengths as the transmission overhead
of binary codes is too large for the short block length regime. \label{fig:energy}}
\end{figure}

Assuming a constant CPU power draw, the delay difference should produce
an equivalent energy reduction. This hypothesis is tested by measuring
the energy consumption for different block lengths, $PLR_{T}=\{0.01,0.001\}$
and $p_{e}=0.05$ (see Figure~\ref{fig:energy}). Each code, including
encoding and decoding, is iterated 100 times to produce visible results
above the background energy when the system is idle so that the energy
consumed by the codes can be isolated using a smoothing average of
30 samples and discarding the samples below a certain threshold (the
middle point between minimum and maximum is typically a good threshold
and thus it has been used for the results here presented). In addition
to the coding operations, each measurement also considers the transmission
overhead in binary codes. The iperf3 tool has been used to transmit
over UDP 100 times the parity packet excess. Figure~\ref{fig:energy}
depicts the average energy consumption of 10 executions for each block
length, including the standard deviation as error bars to ensure that
the achieved differences are statistically significant. The background
energy has been subtracted for the calculation of the total energy
in order to only represent the energy excess when en-/decoding or
transmitting. As expected, the energy consumption follows the same
ratio between codes as the delay in Figure~\ref{fig:Parity-Delay}.
As the block length is reduced, so is the coding energy consumption,
whereas the transmission overhead stays roughly the same. The implication
of such a result is twofold: i) as the transmission overhead accounts
for most of their energy consumption, binary codes that reduce it
in the short block length regime, such as the proposed polar code,
are required, and ii) MDS might potentially be the optimal coding
technique in some cases. Therefore, an energy-aware FEC function is
required for the selection of the coding mechanism that meets the
application's target loss rate and delay with minimum energy consumption.

\subsection{Multicast Performance\label{subsec:Evaluation-Multicast}}

\begin{figure}
\subfloat[$p_{e}=0.01$]{\centering{}\includegraphics[scale=0.35]{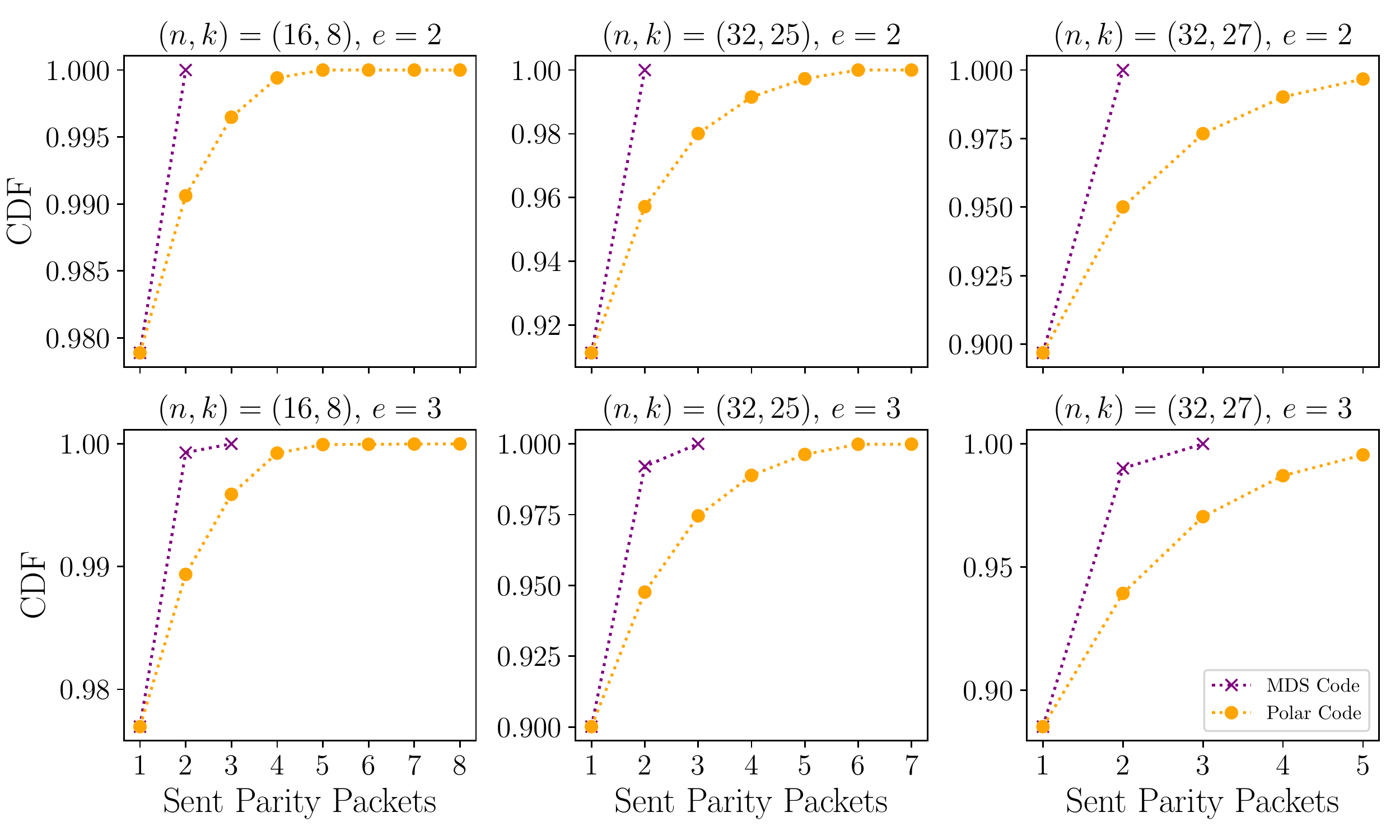}}

\subfloat[$p_{e}=0.05$]{\centering{}\includegraphics[scale=0.35]{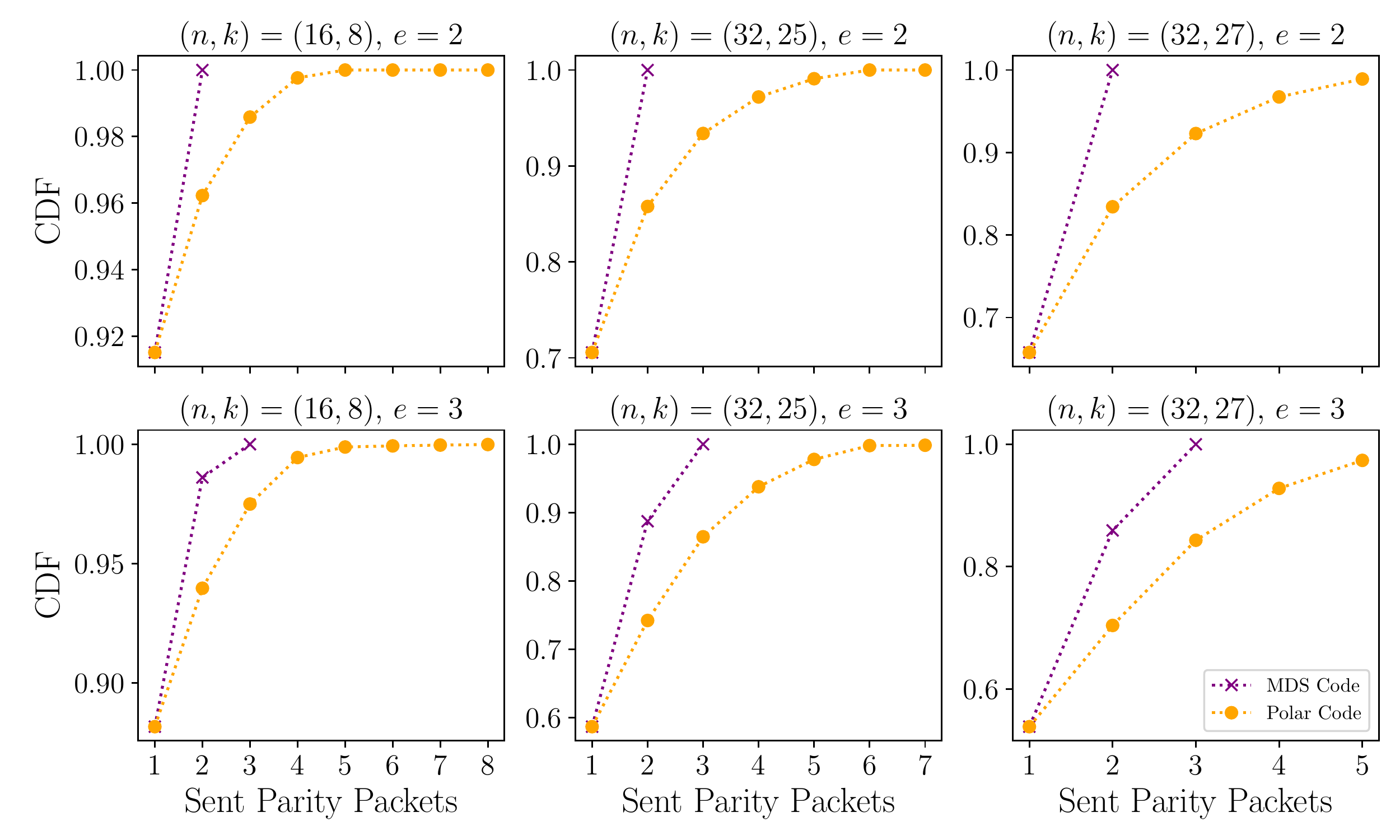}}
\centering{}\caption{Cumulative Distribution Function (CDF) of the erasure recovering probability
in all receivers in a multicast group. The proposed code approaches
the performance of MDS codes.\label{fig:Multicast}}
\end{figure}

For the comparison between MDS codes and the proposed polar code,
a multicast group has been simulated with as many receivers as possible
erasure patterns for $i\in[1,e]$ erasures in a block---$e$ here
is the number of erasures the reference MDS code can correct. After
the transmission of each packet, every receiver performs Gaussian
elimination to check how many erasures it can correct. Since each
erasure pattern has a different probability, each receiver is weighted
differently depending on the number of erasures it originally experienced.
Figure~\ref{fig:Multicast} shows the Cumulative Distribution Function
(CDF) for three different coding configurations and two erasure rates
$p_{e}=\{0.01,0.05\}$. Fountain codes are not depicted because the
amount of correctable patterns always depends on the selected columns
due to their randomness---e.g., after the first parity packet is
transmitted only 50\% of the receivers on average could recover their
losses.

Although the performance of polar code is close to MDS, a closer look
shows that there is still room for improvement. Ordering the columns
according to the Bhattacharyya parameter results in decreasing column
degree. For example, for $n=16$, $k=10$ the degrees of the columns
are $[16,8,8,8,8,4]$. However, the frozen positions do not contribute
any information to the parity packets, so that the column degree is
reduced in practice when there is a one in a frozen bit position.
For the previous example, the degrees are reduced to $[10,6,6,7,7,3]$.
Enforcing decreasing degrees can lead to code-rate dependent reordering
that sacrifices the systematic code construction. Since the degree
difference is small, so is the expected gain of re-sorting and hence
it is not implemented.

\section{Conclusion and Future Work\label{sec:Conclusion}}

The energy aspects of data transmission become fundamental as more
NCSs are deployed on embedded devices. Error coding at the transport
layer can provide timeliness with predictable reliability and we have
shown in this paper that binary codes are an energy-efficient option
when their error handling performance is brought close to MDS codes.
The proposed polar coding outperforms other binary codes in the short-blocklength
regime and approaches MDS codes in multicast, thereby resulting in
an alternative to efficient HARQ for transport layer multicast.

The results presented here show that binary codes fail to achieve
good performance when the required code rates are low. In that case,
MDS codes will potentially be required since they generate lower transmission
overhead. The selection of the most efficient coding technique is
still an open problem and an energy-aware decision considering the
energy consumption for the redundancy computation and transmission
could be an interesting solution.

\section*{Acknowledgments}

This work is supported by the German Research Foundation (DFG) within
SPP 1914 \textquotedblleft Cyber-Physical Networking\textquotedblright{}
under grant 315036956.

\bibliographystyle{IEEEtran}
\bibliography{paper}

\end{document}